\documentclass[preprint,prb,doublespace,showpacs,preprintnumbers,amsmath,amssymb]{revtex4}
\usepackage{graphicx}
\usepackage{dcolumn}
\usepackage{bm}

\begin{document}


\title{
Ion-beam induced 1D to 3D periodic transformation in
  nanostructured multilayers}


\author{S. Bera, B. Satpati, K. Bhattacharjee, P. V. Satyam and B. N. Dev\footnote{email: bhupen@iopb.res.in}}
 \address{Institute of Physics, Sachivalaya Marg,
   Bhubaneswar-751005, India}

\begin{abstract}
Ion-irradiation-induced modifications of a periodic Pt/C multilayer
system containing Fe impurity have been analyzed by transmission
electron microscopy (TEM). The multilayer stack with 16 Pt/C layer
pairs (period 4.23 nm) was fabricated on a glass substrate. A 2 MeV
Au$^{2+}$ ion beam was rastered on the sample to obtain uniformly
irradiated strips with fluences from 1$\times10^{14}$ to
1$\times10^{15}$ $ions/cm^2$. Ion-irradiation has been found to cause
preferential migration of Fe towards Pt layers [Nucl. Instr. Methods
Phys. Res. B212 (2003) 530]. Cross-sectional transmission electron
microscopy (XTEM) shows considerable atomic redistribution for
irradiation at the highest ion fluence  (1$\times10^{15}$
$ions/cm^2$). Individual entities in this structure is like a
cluster. Periodic multilayers have periodicity only in the direction
normal to the multilayer surface. However, Fourier transform of the
XTEM images of the sample irradiated at the highest-fluence shows new off-normal Fourier components
of superlattice periodicities arising due to ion irradiation. With a proper understanding of this phenomenon it may be possible to fabricate three dimensional periodic structures of nanoclusters.

\end{abstract}
\maketitle

\section{INTRODUCTION}
Multilayer structures, nanostructured in one dimension, have unique
structural\cite{ref[1]}, magnetic\cite{ref[2]} and electronic\cite{ref[3]} properties with a wide range of
applications. Artificial multilayers are metastable due to their high
content of interfaces as well as their metastable microstructure
arising from specific conditions of layer deposition. Therefore the
stability and modification of multilayers under thermal and ion
irradiation treatment have important consequences for their
appropriate applications. A magnetic multilayer contains alternating magnetic and nonmagnetic layers. Ion-beam irradiation of magnetic multilayers
have been found to show a spin-orientation transition, indicating the
suitability of such systems for patterned ultrahigh density recording
media\cite{ref[4]}. Structural analysis has shown that in periodic multilayers
ion-beam irradiation causes mixing, interface broadening and period
dilation\cite{ref[5]}. All these changes may be responsible for the
spin-orientatation transition in magnetic multilayers. However, a proper correlation of properties with these parameters is still lacking. In a magnetic multilayer, a small concentration of magnetic impurity in the nonmagnetic layer can drastically change magnetic coupling and magnetoresistance\cite{ref[6]}. As ion beam can introduce such magnetic impurities into the nonmagnetic layers by ion-induced displacements of atoms from the magnetic layers into the nonmagnetic layer\cite{ref[5]}, ion irradiation could possibly be used to tune magnetic coupling and modify magnetoresistance. In  Pt/C
multilayers with Fe impurities [ hereafter denoted as Pt/C(Fe) ], ion-irradiation has been found to cause preferential migration of Fe towards Pt layers\cite{ref[7]}. Considering
that Fe-Pt forms magnetic alloys, this phenomenon raises the
possibility of fabrication of magnetic clusters with the nonmagnetic
layers of the multilayer providing the non-magnetic matrix to isolate
the magnetic particles. We have investigated the aspect of ion-beam induced cluster formation in Pt/C(Fe) multilayers.

We have carried out transmission electron diffration (TED) and microscopy (TEM) studies on virgin
and ion-irradiated Pt/C(Fe) multilayer systems to investigate the morphological modification due to ion bombardment. One-dimensional periodicity of the multilayer has been found to evolve into a quasi-three-dimensional periodicity upon ion irradiation. These aspects of ion beam modification are presented here.

\section{ Experimental}
Pt/C multilayers were fabricated on float glass substrates, by ion
beam sputtering, at a low Argon pressure of 0.1 mbar. Fe impurity was
introduced in the multilayers during growth. The expected Fe
concentration in both the Pt-layers and C-layers was slightly over 10 $at$
\%. The Pt/C(Fe) multilayer sample used in this study, was prepared at
Nagoya University. The sample specifications are: N = 16 (the number
of layer-pairs in the multilayer stack), $d$ = 4.23 nm (multilayer
period, i.e. the thickness of a Pt/C layer-pair),  $\Gamma$  = 0.38 (
ratio of Pt layer thickness to $d$ ). The total thickness of the
multilayer stack is about 68 nm. Different parts of a large sample
(30$\times70$)mm$^{2}$ were irradiated with 2 MeV Au$^{2+}$ ions by
rastering the ion beam on (30$\times5$)mm$^{2}$ strips at various
fluences ($ions/cm^2$) [A(virgin), B(1$\times10^{14}$),
C(3$\times10^{14}$), D(5$\times10^{14}$), F(1$\times10^{15}$)] at
Institute of Physics, Bhubaneswar. The range of 2 MeV Au ions in such
Pt/C multilayers
sample is about 270 nm. Thus the implanted Au ions are buried deep
into the glass substrate well past the multilayer stack, which is about 68 nm thick. The virgin and irradiated strips were analyzed
with X-ray standing wave and X-ray reflectivity experiments. X-ray
standing wave and reflectivity experiments were carried out at Hamburg
Synchrotron Radiation Laboratory (HASYLAB at DESY), Hamburg, at the
ROEMO-I beamline using 14.0 keV monochromatized X-rays. Details of
these results will be presented elsewhere. The TEM measurements on a virgin sample (A) and on the sample irradiated at the highest fluence (F) were carried out with 200 keV electrons using a JEOL 2010(UHR) electron microscope at our Institute. Here we mainly present the results of our TEM measurements.

\section{ Results and discussions}
X-ray reflectivity scans showing the first to the fourth order Bragg peaks (Fig. 1) from the virgin and the irradiated (fluence: 1$\times10^{15}$ $ions/cm^2$) Pt/C(Fe) samples indicate that the periodic structure in the surface-normal direction of the multilayer is preserved in the irradiated sample. The first-order Bragg peaks are shown in the linear scale in the inset of Fig. 1. For the irradiated sample the intensity of the Bragg peak is reduced and the position of the Bragg peak indicates that there is a contraction of the multilayer period compared to the virgin sample\cite{ref[8]}. In fact the Bragg peaks of all orders are systematically shifted to higher angles. We do not get additional information about the morphological changes. A cross-sectional TEM (XTEM) image of the virgin sample (Fig. 2) shows the periodic multilayer structure with the top and the bottom Pt layer thicker than other Pt layers. Fourier transform (FT) of the marked square region of the XTEM image in Fig. 2, presented in the inset (a), shows this periodicity in one dimension in the direction normal to the multilayer surface. The diffraction pattern from the multilayer displaying the 1D periodicity is shown in the inset (b). This one-dimensional (1D) periodic pattern is expected from the structure seen in Fig. 2 and its FT. In order to probe the crystallinity of the constituent layers, a selected area diffraction pattern from the multilayer (virgin sample) is shown in Fig. 3. From the diffraction pattern it appears that the Pt layers are polycrystalline, while the C layers are amorphous. The indexed spots are from polycrystalline Pt. No diffraction features from C are observed. The amorphous nature of C is also evident from the higher resolution XTEM image in Fig. 4, where lattice images of Pt crystallites are seen. A XTEM image of the irradiated multilayer shows atomic redistribution and cluster formation (Fig. 5). The FT, taken from the marked square region in Fig. 5 is shown in the inset (a). In addition to the spots in the surface-normal direction, evolution of new spots in the FT shows the development of new periodicities within the irradiated multilayer. Compared to those in Fig. 2, new Fourier components in the off-normal direction are seen in Fig. 5. Apparently ion irradiation has produced a three dimensional quasi-periodic structure from a structure which was periodic only in one-dimension. The continuous layers in the virgin sample have evolved into clusters upon ion irradiation.

Let us try to understand why clusters are formed due to ion irradiation. It was shown that in a heterogeneous system all kinds of mixing processes associated with ion bombardment, including ballistic and chemically guided atomic movements, can be incorporated in a diffusion equation\cite{ref[9],ref[10],ref[11]}. The sign of the effective diffusion co-efficient can be either positive or negative. Positive coefficient leads to mixing between components and negative co-efficient leads to phase separation. The sign of diffusion co-efficient is determined by the heat of mixing of the components. The sign of a factor $F$ giving the sign of the diffusion coeffecient is determined by
\begin{equation}
F = 1 -  \frac{2\Delta H_m^{1:1}}{k_BT}\frac{p}{p+1}  
\end{equation}
as given by Miotello and Kelly\cite{ref[9],ref[10]} ignoring the concept of thermal spikes due to energetic penetrating ions. Another expression given by Cheng\cite{ref[11]} incorporating atomic movements in thermal spikes is
\begin{equation}
 F = 1 - 27.4\frac{\Delta H_m^{1:1}}{\left|\Delta H_{coh}\right| } 
\end{equation}
In Eqs. (1) and (2) $\Delta H_m^{1:1}$ is the heat of mixing of the alloy containing 50 $at$ \% of each component, $\left|\Delta H_{coh}\right|$ is the absolute value of the average cohesive energy of the system, $T$ is the absolute temparature, $k_B$ is the Boltzmann's constant and $p$ is the ratio of the diffusion coefficients describing the chemically guided and the random atomic relocation. The value of $p$ is found to increase rapidly as temparature is raised. Its value at room temperature, at which our experiments were carried out, is around $p \approx 0.1$, independent of the materials of the pair to be irradiated\cite{ref[9]}.

On the basis of Miedema's ``macroscopic atom model"\cite{ref[12]}, we have calculated the change of enthalpy  $\Delta H_m$ of Pt-C mixing. The enthalpy change can be expressed as\\
\begin{eqnarray}
\Delta H_m =&&\frac{2Pf(c^s)(c_{Pt}V_{Pt}^{2/3} + c_CV_C^{2/3})}{(n_{ws}^{Pt})^{-1/3}+ (n_{ws}^C)^{-1/3}}\biggl(-(\Delta\phi^\ast)^2 + \frac{Q}{P}(\Delta n_{ws}^{1/3})^2 - \frac{R}{P}\biggr) \nonumber\\
&&+ c_C[\Delta H (C_{element} \rightarrow C_{metal})]
\end{eqnarray}
where
\begin{eqnarray*}
&f(c^s) = c_{Pt}^sc_C^s\\
&c_{Pt}^s = c_{Pt}V_{Pt}^{2/3}/(c_{Pt}V_{Pt}^{2/3} + c_CV_C^{2/3}) \\
&c_C^s = c_CV_C^{2/3}/(c_{Pt}V_{Pt}^{2/3} + c_CV_C^{2/3})\\
&P = 12.3,  \frac{Q}{P} = 9.4 V^2 (d.u.)^{-2/3},  \frac{R}{P} = 2.1 V^2
\end{eqnarray*}
Here $c_{Pt}$ and $c_C$ are atomic concentrations of Pt and C respectively. $V_{Pt}$ and $V_C$ are the molar volumes of Pt and C respectively. $c^s$ is the surface concentration. $\phi^\ast$ is a modified work function and $n_{ws}$ the electron density at the first Wigner-Seitz boundary. $P$, $Q$ and $R$ are constants. $\Delta H_m$ is expressed in kilojoules per mole, $\phi^\ast$ in volts, $n_{ws}$ in density units (d.u.) and $V^{2/3}$ in square centimetres. In equation (3), the first term represents the charge transfer between Pt and C. The second term represents the discontinuity in electron density at the Wigner-Seitz boundary of Pt and C atoms. The third term represents a hybridization contribution. Both the first and the third terms, which are negative, favor a tendency for compound formation, while the second term favors a tendency for phase separation. The last term is a transformation energy that accounts for the enthalpy difference between elementary and metallic carbon. According to the theory of Miedema\cite{ref[17]}, alloys of transition metals with C can be treated as the other alloys of transition metals with polyvalent non-transition metals. The only difference is to introduce an additional positive contribution to account for the enthalpy difference between carbon in diomond structure and a more conventional metallic structure. For carbon, the corresponding transformation energy equals 180 kJ/mole\cite{ref[14]}.

The cohesive energy of a Pt-C alloy $\left|\Delta H_{coh}\right|$ can be approximated using regular solution theory\cite{ref[13]} by\\
\begin{equation}
\left|\Delta H_{coh}\right| \cong (c_{Pt} H_{Pt}^0 + c_C H_C^0) + \Delta H_m
\end{equation}
where $H^0$ is the cohesive energy of individual element.  $\Delta H_m$ is the regular solution heat of mixing. The values of $\phi^\ast$, $n_{ws}^{1/3}$, $V^{2/3}$ and $H^0$ for C, Pt, and Fe are listed in Table I.
Using  equations (1), (2), (3) and (4), we have calculated  $\Delta H_m^{1:1}$, $\left|\Delta H_{coh}^{1:1}\right|$ and $F$ for Pt-C, Pt-Fe and Fe-C systems. Calculated values are listed in Table II. The value of $F$ in column four have been calculated for T = 300K.
For
the Pt-C system, a negative diffusion co-efficient corresponds to
phase separation in Pt-C. The large negative $F$ factor indicates that the driving force for separation of the components (Pt and C) in the ion-bombarded sample is very pronounced. This phase seperation could be the reason for formation of Pt nanoparticles surrounded by C. This point is further discussed in the following paragraph. The positive value of $F$ for the Pt-Fe system indicates that mixing is favored. The negative and positive values of $F$ for the Fe-C and the Fe-Pt, respectively, also explain the preferential migration of Fe from C- to Pt-layers as observed in our earlier studies\cite{ref[7]}.

As observed here, other recent HRTEM studies on Pt/C multilayers to explore atomic scale
details of the interfaces and individual layers have shown that the C
layers are amorphous and the Pt layers are polycrystalline\cite{ref[20]}. The Pt
grain size perpendicular to the layers was about the layer thickness
while the grain size along the layers varied in the range up to 10
nm\cite{ref[20]}. The interface profiles were found to be in agreement with the previous X-ray
analysis by Ghose and Dev\cite{ref[21]} showing that Pt-on-C interface is broader than the
C-on-Pt interface - apparently a consequence of higher surface free energy of Pt
compared to C. The nanocrystalline nature of the Pt layers, together
with the ion irradiation effect explained below, is
helpful in understanding the morphology and the evolution of new
periodicities seen in Fig. 5. Ion beam induced atomic displacements in
a Pt/C multilayer, as shown in the simulation in Ref. 5, causes
mixing at the interface. That is, C is introduced into the Pt layers and vice
versa. These atomic redistributions are the results of an athermal process. However,
radiation-enhanced-diffusion (especially at higher fluences as the case here) and chemical drive may influence this
distribution. For the Pt-C case the heat of mixing yields a negative
value of $F$ indicating that phase separation would take place in this
system. Thus C, incorporated into Pt by ion beam, would diffuse out of
Pt by a chemically driven process. This C can accumulate in the grain
boundary region between Pt nanocrystalline grains within the Pt
layers leading to formation of isolated Pt nanoparticles. In our previous study ion fluence was lower\cite{ref[5]}. At this fluence the Pt/C layers preserve their identity as continuous layers. Compared to this the ion
fluence in the present work is an order of magnitude higher. This does not necessarily produce
 a stronger ion-beam induced atomic displacement and mixing at the end of the irradiation process. Besides ballistic mixing, the role of the chemically guided diffusion is to be taken into account. This process is well documented in diffusional relaxation of ion-bombarded systems\cite{ref[22],ref[23]}. Inclusion of this factor [Eqs. (1) and (2)]
leads to phase separation for the Pt/C system leading to Pt
nanoparticales surrounded by C.

It was previously observed that Fe
impurity atoms in the Pt/C multilayers, upon ion irradiation, migrate
preferentially from C layers to Pt layers\cite{ref[7]}. This was explained by
comparing Fe-Pt\cite{ref[24]} and Fe-C\cite{ref[25]} phase diagrams. In conjunction with this
behavior, the structure in Fig. 5, indicates that the Pt
nanoparticles contain Fe and formation of FePt nanoparticles cannot be ruled out. FePt is an interesting and useful material
because in the face centered tetragonal $L1_0$ chemically ordered phase they
exhibit
a very high magnetic anisotropy and gaint magneto-optical effects.
In order to search for FePt formation, we carried out TEM studies at higher resolution on various parts of sample represented in Fig. 5. One XTEM image from these studies is shown in Fig. 6, where we find two sets of lattice images, one corresponding to d = 0.218 $\pm$ 0.005 nm and another corresponding to d = 0.251 $\pm$ 0.007 nm. d = 0.218 nm is closer to the (111) planar spacing of bulk Pt. The lattice spacing of d = 0.251 nm does not match with any available d spacing of either Pt or Fe structures. However, it is close to the (110) planar spacing of bulk FePt. We believe, this provides an evidence for FePt nanoparticle formation.
With the 
method presented here, one may visualize the possibility of fabrication of a three-dimensional (3D) quasi-periodic lattice of FePt magnetic nanoparticles with an appropriate Fe concentration. An ordered array of such nanoparticles is expected to show unusual properties due to interparticle coupling.

\section{Conclusions}
We have carried out irradiation of Pt/C multilayers, containing a
small  amount of Fe impurity, with 2 MeV Au$^{2+}$ ions at different
ion fluences. Cross sectional transmission electron microscopy (XTEM) of
ion-irradiated (fluence 1$\times10^{15}$ $ions/cm^2$) Pt/C(Fe)
multilayers shows the formation of clusters within the layers. Fourier
transform of the XTEM micrographs shows new Fourier components representing
new superlattice periodicities arising due to ion irradiation. A 1D-periodic multilayer appears to undergo a transformation to a 3D quasi periodic structure upon ion irradiation. With a
proper understanding of this phenomenon it may be possible to
fabricate three dimensional periodic structures of nanoclusters by ion
irradiation. Considering the preferential Fe migration to Pt layers and the
possible magnetic phases of Fe-Pt alloys, our observation raises 
the possibility of fabrication of ion-beam-induced magnetic nanocluster lattices.\\

\section{Acknowledgment}
We thank Prof. K. Yamashita for providing the virgin Pt/C(Fe) sample and Dr. G. Kuri for the help in X-ray measurements.

\newpage
FIGURE CAPTION\\

\vspace*{0.2in}

FIG. 1:  Experimental X-ray reflectivity data from the virgin (circle) and  the
ion-irradiated (square) multilayer samples. Data for the irradiated sample are
vertically shifted down by three orders of magnitude for clarity. The first-order Bragg
peaks are shown in the linear scale in the inset.\\

\vspace*{0.4in}

FIG. 2:  A cross-sectional TEM image of the virgin Pt/C(Fe) multilayer. Fourier transform of the marked square region is shown in the inset (a). Diffraction pattern due to the periodic multilayer stack is shown in the inset (b).\\
\vspace*{0.4in}

FIG. 3:  A selected area transmission electron diffraction pattern from the multilayer. The indexed spots are from polycrystalline Pt.\\
\vspace*{0.4in}

Fig. 4: A HRTEM image from the virgin Pt/C(Fe) multilayer.\\

\vspace*{0.4in}

FIG. 5: A cross-sectional TEM image of the irradiated Pt/C(Fe) multilayer. Fourier transform of the marked square region of the TEM image is shown in the inset (a). Inverse Fourier transform of inset (a) is shown in the inset (b).\\

\vspace*{0.4in}
Fig. 6: A HRTEM image from the irradiated sample. The region with the planar spacing of 0.251 nm is likely to be a FePt region.\\

\newpage
\begin{table}
\caption{The values of $\phi^\ast$, $n_{ws}^{1/3}$, $V_m^{2/3}$ and $H^0$ for calculating $\Delta H_m$.}
\begin{center}
\begin{tabular}{lllllllllll}
\hline
&&&$\phi^\ast$&&&$n_{ws}^{1/3}$&&$V_m^{2/3}$&&$H^0$\\
&&&$(V)$&&&$((d.u.)^{1/3})$&&$(cm^{2})$&&$(kJ/mole)$\\
&&&$[15-18]$&&&$[15-18]$&&$[19]$&&$[19]$\\
\hline
C&&&6.20&&&1.90&&2.26&&531\\
Pt&&&5.65&&&1.78&&4.36&&564\\
Fe&&&4.93&&&1.77&&3.70&&413\\
\hline
\end{tabular}
\end{center}
\end{table}
\vspace*{5in}
\newpage
\begin{table*}
\caption{Values of heat of mixing, cohesive energy and the $F$ factor determining the sign of the effective diffusion coefficient for the Pt-C, Pt-Fe, Fe-C systems.}
\begin{center}
\begin{tabular}{lllllllllllll}
\hline
$System$&&&$\Delta H_m^{1:1}$&&&$\left|\Delta H_{coh}^{1:1}\right|$&&&$F$&&&$F$\\
&&&$(kJ/mole)$&&&$(kJ/mole)$&&&$[9,10]$&&&$[11]$\\
\hline
Pt-C&&&52.7&&&600.0&&&-2.8&&&-1.4\\
Pt-Fe&&&-12.1&&&476.4&&&1.8&&&1.7\\
Fe-C&&&35.3&&&507.3&&&-1.5&&&-0.9\\
\hline
\end{tabular}
\end{center}
\end{table*}



\begin{thebibliography}{}
\bibitem{ref[8]} In our earlier irradiation experiment at a lower fluence we observed a period dilation [ref. 5]. Here also we have observed (not presented here) period dilation for an ion fluence of 3$\times10^{14}$ $ions/cm^2$.
\end{thebibliography}
\end{document}